\documentclass[%
 reprint,
 amsmath,amssymb,
 aps,
]{revtex4-1}

\usepackage{graphicx}
\usepackage{dcolumn}
\usepackage{amsmath}
\usepackage{bm}
\usepackage{xcolor}
\usepackage{array}
\usepackage{wrapfig}
\usepackage{appendix}
\usepackage[T1]{fontenc}
\usepackage[utf8]{inputenc}
\usepackage[english]{babel}

\AtBeginDocument{
  \def\selectlanguage#1{}
}

\begin{document}

\title{Possible fractal nature of accretion flows in MAD and SANE simulations: Implications to GRS 1915+105}

\author{Srishty Aggarwal$^1$}
\email{srishtya@iisc.ac.in}

\author{Rohan Raha$^1$}
\email{raharohan@iisc.ac.in}

\author{Mayank Pathak$^2$}
\email{mayankpathak@iisc.ac.in}

\author{Banibrata Mukhopadhyay$^{1,2}$}
\email{bm@iisc.ac.in}

\affiliation{$^1$Department of Physics, Indian Institute of Science, Bangalore 560012, India\\
$^2$Joint Astronomy Programme, Department of Physics, Indian Institute of Science, Bengaluru 560012, India}


\begin{abstract}
The general relativistic magnetohydrodynamic (GRMHD) simulations are widely used to study accretion disk and jet dynamics around a black hole. Despite strong observational evidences for intrinsically nonlinear behavior, the interpretations of GRMHD simulation results, more precisely the underlying timeseries, have not been well-explored by nonlinear timeseries analysis. In this work, we characterize the jet and disk dynamics of different GRMHD simulated flows using the nonlinear timeseries analysis. As diagnostic tools, we consider Higuchi fractal dimension (HFD), Hurst Index ($H$) and spectral slope. We implement them for two model disk frameworks: magnetically arrested disk (MAD) and standard and normal evolution (SANE), across a range of black hole spins with the Kerr parameter spanning from -0.9375 to 0.9375. We simulate the disk/jet systems by two well-documented codes: HARMPI and BHAC, and obtain, respectively, low and high temporally resolved timeseries data. For both jet and disk dynamics, MADs are characterized by higher HFD, lower $H$ and flatter spectral slopes than SANEs.  High HFD in MAD could be due to its intermittent variability and indicates that it has lesser long-range temporal correlations than SANE. Moreover, HFD in MAD decreases with spin magnitude owing to increase in collimated, hence ordered, jets. However, in SANE, it increases with spin for positive ones due to interplay of winds and jets. Extending our analysis to observations, we attempt to segregate the classes of black hole: GRS 1915+105, into MAD- and SANE-like clusters based on their spectral properties extracted from X-ray data. The mean HFD of MAD-like cluster is higher than SANE-like cluster, thus, corroborating with the simulation results. This suggests that SANE systems are more correlated than MAD. This further corroborates with MAD's eruption with strong magnetic fields. Our work highlights the role of nonlinear timeseries analysis to understand the underlying dynamics of accretion flows and their connection to magnetic regulation.
\end{abstract}
\maketitle

%

\section{Introduction}
Accretion onto compact objects is among the most efficient mechanisms of energy release in the universe, powering systems that range from X-ray binaries to active galactic nuclei (AGNs) and quasars. The temporal variability of accretion flows carries direct information about the underlying transport processes, turbulent cascades, and magnetic field dynamics that govern angular momentum transfer and energy dissipation. 
Traditionally, accretion variability, particularly based on observed data, has been characterized using linear statistical measures, primarily power spectral density (PSD) and 
auto-correlation functions \cite{Nowak1999, vanderKlis2006}. These approaches successfully revealed ubiquitous red-noise spectra, quasi-periodic oscillations (QPOs), and broadband turbulence in both stellar-mass and supermassive black hole systems \cite{vanderKlis2006, McHardy2006}. However, 
they are limited in their ability to quantify long-range memory, scale-invariant complexity, and deterministic chaos.

On the other hand, nonlinear techniques have been used to uncover rich astrophysical dynamics. Correlation dimension and Lyapunov exponent analyses of accretion sources such as GRS 1915+105 indicated signatures of low-dimensional chaos, challenging purely stochastic interpretations of disk variability \cite{misra_chaotic_2004, harikrishnan_non-subjective_2006}.

One of the most common methods for fractal analysis, which also has been used for astrophysical signals, has been the correlation dimension by Grassberger and Procacia \cite{grassberger_measuring_1983}, which infers the dimensionality of the underlying dynamical attractor through phase-space reconstruction. However, its reliable estimation requires long, high-quality, and approximately stationary timeseries with  $\gtrsim10^4$
 data points, and is highly sensitive to observational noise, finite-size effects, and the choice of embedding parameters \cite{theiler_spurious_1986, KantzSchreiber2004}. These constraints are rarely satisfied in astrophysical contexts for both observational data and numerical simulations, where timeseries are often of limited length and intrinsically non-stationary due to evolving physical conditions. In this regard, the Higuchi Fractal Dimension (HFD) provides a robust and computationally efficient alternative. The Higuchi's method directly estimates the fractal dimension from the timeseries without phase-space reconstruction, remains reliable for short and non-stationary signals, and has been shown to be robust in the presence of noise \cite{higuchi_approach_1988,accardo_use_1997}. Although introduced in 1988 \cite{higuchi_approach_1988}, its use in astrophysical applications has remained extremely limited, largely restricted to solar and space-physics studies, such as analyses of solar radio emission, magnetic field indices, and solar wind variability \cite{watari_fractal_1996, bhatt_variability_2018}.   Moreover, fractal dimension measurements of solar active regions demonstrated that magnetic complexity plays a decisive role in regulating flare productivity and energy release \cite{mcateer_statistics_2005}, while multifractal analyses of stellar photometric variability revealed scale-dependent energy transport and mode coupling in stellar envelopes \cite{deFreitas2019}. 

Another important nonlinear diagnostic is the Hurst index ($H$), which characterizes long-range temporal correlations and memory in timeseries \cite{hurst_long-term_1951}. It has been applied on artificial data and simulated sunspot magnetic field maps, as well as to observations obtained with NASA’s Marshall Space Flight Center vector magnetograph, and found to be sensitive to large-scale magnetic structures in its two-dimensional images \cite{adams_study_1997}. More recently, Hurst-based analyses have been employed in solar-cycle studies, including attempts at early prediction of 25th Solar Cycle \cite{singh_early_2017}. Beyond solar physics, it has also been applied to gamma-ray burst (GRB) lightcurves, where it has been correlated with multiple observational parameters through regression and correlation analyses, providing insight into the temporal organization of GRB emission processes \cite{guan_hurst_2025}.
Together, these works have demonstrated that fractal and memory-based measures provide physically meaningful diagnostics of astrophysical systems, motivating their application to accretion disk variability.

The advent of high-resolution general relativistic magnetohydrodynamic (GRMHD) simulations over the last decade has revolutionized our understanding of accretion physics, enabling direct modeling of black hole disks, jets, and radiation. These simulations have revealed two distinct accretion regimes: the magnetically arrested disk (MAD) state, characterized by the saturation of large-scale vertical magnetic flux near the black hole horizon and powerful jet production \cite{narayan_magnetically_2003, tchekhovskoy_efficient_2011}, and the standard and normal evolution (SANE) state, in which the magnetic field remains dynamically subdominant, but turbulence is driven primarily by the magnetorotational instability (MRI) \cite{narayan_grmhd_2012} based on weak fields. With the advent of very long baseline interferometry (VLBI) techniques allowing for microarcsecond angular resolutions, the role of GRMHD simulations has become more relevant. Collaborations like Event Horizon Telescope (EHT) have used GRMHD simulations for interpreting accretion and black hole properties of supermassive black holes like M87*. The accretion flow around M87* has been understood to be a MAD system \citep{tsv98,cyl23}. While the MAD systems are understood to be accretion flows with jets, e.g. certain temporal classes of GRS~1915+105 \cite{Belloni2000}, the SANE systems are generally understood to be the systems with weak or no jets (but may have winds), e.g., certain other classes of GRS~1915+105 \cite{Belloni2000,rohan2026}. 

Despite detailed studies of GRMHD simulations \citep{raha2023,rohan2026}, particularly around black holes, with powerful jets have been conducted to  understand turbulence and reconnection-driven flaring \cite{Hawley2001, McKinney2012, bart19}, nonlinear tools have not been employed to quantify their memory effects, fractal geometry of the signal, or 
scale-dependent persistence. Crucially, no systematic nonlinear comparison of the MAD and SANE accretion flows enlightening the underlying possible fractal nature currently exists in the literature. In this work, we employ different nonlinear techniques, namely HFD and $H$, along with spectral slope to characterize the simulated jet and disk dynamics for MAD and SANE profiles.

Additionally, we attempt to validate the simulation results with the observational data of the black hole source GRS~1915+105. This is one of the most extensively studied black hole X-ray binaries, notable for its extreme variability of timescales ranging from seconds to months as well as near-Eddington luminosity. Their lightcurves, obtained using Rossi X-ray Timing Explorer (RXTE), have been grouped into 12 distinct temporal classes by Belloni et al. \cite{Belloni2000}. These classes exhibit complex, aperiodic, and often highly 
non-stationary lightcurves. Hence, they have been analyzed using various nonlinear techniques, like correlation dimension, correlation entropy, singular-value decomposition and multifractal spectrum \cite{misra_chaotic_2004, harikrishnan_non-subjective_2006, harikrishnan_nonlinear_2011,pradeep_measuring_2023}. Mukhopadhyay and collaborators even identified four different accretion modes by correlating the nonlinear properties of these classes based on timeseries with their spectral states \cite{adegoke_correlating_2018}. In the present work, we classify the 12 classes of GRS~1915+105 into two clusters based on their spectral properties and identify the clusters to be MAD-- and SANE--like. We then calculate HFD for their lightcurves and compare it on average for MAD-- and SANE--like clusters to check consistency with simulation results.

The paper is organized as follows. The details about the simulation setup are given in Sec. \ref{Sec_simsetup}, followed by the description of analysis techniques in Sec. \ref{Sec_analysesTEch}. 
The results based on simulations are provided in Sec. \ref{Sec_Results} and their observational implications and connections are composed in Sec. \ref{Sec:GRS}. We conclude in Sec. \ref{Sec_Conclusion} with a discussion of future implications of this work.

\section{\label{Sec_simsetup} Simulation Setup}
We study the spatio-temporal evolution of a magnetized accretion flow around a black hole using two different simulation codes, HARMPI and BHAC. The accretion flow is initiated by using the Fishbone-Moncrief (FM) \citep{fm} torus solution. 

The equations solved by the codes are:
\begin{equation}
\begin{aligned}
    \nabla_\mu(\rho u^\mu)&=0,\\
    \nabla_\mu T^{\mu\nu}&=0,\\
    \nabla_{\mu}\hspace{0.01in}^*F^{\mu\nu}&=0,   
\end{aligned}
\end{equation}
where $u^{\mu}$ is the four-velocity, $T^{\mu\nu}$ is the stress-energy tensor and $^{*}F^{\mu\nu}$ is the dual Faraday tensor. Here, $\mu$ and $\nu$ are spacetime indices such as $t,r,\theta,\phi$.

For non-dimensional evolution of the flow equations, we use geometric units, i.e., $GM=c=1$, hence $r_{g}=GM/c^2=1$ and the light crossing time, $t_g\equiv \rm{r_g/c} = 1$, in our simulations.
The initial magnetic field geometry in the simulation domain is defined by two formalisms, namely SANE and MAD. Their corresponding magnetic vector potentials are as follows \citep{rohan2026}:
\begin{enumerate}
    \item MAD: $A_\phi=\max(\rho/\rho_0\exp(-r/400)(r/r_{\mathrm{in}})^3\sin^3\theta-0.2,0)$,
    \item SANE: $A_{\phi}=\max(\rho/\rho_0-0.2,0)$,
\end{enumerate}
where $\rho_0$ is the maximum density in the initial torus, set at $r=41$, while $r_{\mathrm{in}}=20$ is the inner edge of the FM torus.
The initial field strength is set by defining the initial plasma-$\beta$ parameter to be $100$ (plasma-$\beta=p_{\mathrm{gas}}/p_{\mathrm{mag}}=p_{\mathrm{gas}}/(b^2/2)$, where $p_{\mathrm{gas}}$ is the gas pressure and $b^2=b^\mu b_\mu$ is the norm of the four-magnetic field).

We study the temporal behavior of the following quantities at the event horizon of the central black hole:
\begin{enumerate}
    \item Outflow Efficiency: $\eta=P_c/\dot{M}$
    \item Normalized magnetic flux: \\
    $\phi=\frac{\sqrt{4\pi}}{2\sqrt{\dot{M}}}\int\sqrt{-g}B^r\mathrm{d}\theta \mathrm{d}\phi$,
\end{enumerate}
where $\dot{M}$ is the accretion rate, defined by $\dot{M} = -\int \sqrt{-g}\:\rho u^r  \, \mathrm{d}\theta \mathrm{d}\phi$ and $P_c$ is the outflow power defined as $P_c=\Dot{M}-\Dot{E}$, $\Dot{E}=\int \sqrt{-g}T^r_t\mathrm{d}\theta \mathrm{d}\phi$ and $T^{\mu}_{\nu}=(\rho+p+u_g+b^2)u^{\mu}u_{\nu}+(p+b^2/2)\delta^\mu_\nu-b^{\mu}b_{\nu}$ is the stress-energy tensor.

We analyse the simulated data for both positive and negative spins: $0.9375,\:0.5,\:0.3,\:0,\:-0.3,\:-0.5,\:\mathrm{and}\:-0.9375$, representing prograde and retrograde motion respectively.  We have evolved 2.5-dimensional (flow evolution in the $r$ and $\theta$ direction with axi-symmetry in the $\phi$ direction) simulations to $30000\:\mathrm{r_g/c}$ timesteps. We perform all the analyses for last $5000$ time steps, i.e. time period of $25000-30000\:\rm{r_g/c}$.
Next, we provide the details of simulation codes. 

\subsection{HARMPI}
HARMPI code is a parallel implementation of a flux-conservative, high-resolution shock-capturing GRMHD code that originated from the 2D High-Accuracy Relativistic Magnetohydrodynamics (HARM) code \citep{Gammie2003}. Variables are zone-centered, and the divergence constraint is enforced using the Flux-CT constrained transport algorithm of \citet{Toth2000}. The code uses piecewise parabolic interpolation for the reconstruction of primitive variables at cell faces and the electric field at the cell edges for the constrained transport scheme, as well as new schemes for ensuring a physical set of primitive variables is always recovered. HARMPI was also written to be agnostic
to coordinate and spacetime choices, making it in a sense generally covariant.  Fluxes are evaluated using the local Lax-Friedrichs (LLF) method. Parallelization is
achieved with a hybrid MPI/OpenMP domain decomposition scheme. HARMPI has demonstrated convergence at second order on a suite of problems in Minkowski and Kerr spacetimes \citep{Gammie2003}.  A public release version of the code can be obtained from \url{https://github.com/atchekho/harmpi}.
We have employed a spatial resolution of $448\times192\times1$ and timestep $10\:\rm{r_g/c}$.

\subsection{BHAC}
The Black Hole Accretion Code (BHAC) is an MPI-Adaptive Mesh Refinement Versatile Advection Code (AMRVAC) based GRMHD simulation code \citep{kepp,porth14,xia18}. This publicly available code is capable of solving 2D and 3D GRMHD systems by utilities the adaptive mesh refinement (AMR) approach based on an oct-tree block-based approach. This strategy leads to efficient allocation of computational resources and in achieving high spatial resolution wherever and whenever needed. The evolution of the GRMHD equations is carried out using a variety of second-order finite-volume schemes, together with magnetic field update methods that incorporate various divergence cleaning techniques like generalized Lagrangian multiplier and flux-interpolated constraint transport \citep{dedner02}. The code is equipped with various modular setups of standard GRMHD systems in ideal and resistive MHD frameworks \citep{bart19}. The publicly available version of the code is hosted at \url{https://bhac.science}. The spatial resolution is $384\times192\times1$ and the timestep is 0.1 $\rm{r_g/c}$.

\section{\label{Sec_analysesTEch}Analysis Techniques}
Primarily, we use HFD to characterize different simulated timeseries. We also compute $H$ and slope of PSD to increase the reliability of the results.

\subsection{Higuchi Fractal Dimension}
The Higuchi Fractal Dimension quantifies the degree of self-similarity and geometric complexity of a 
one-dimensional timeseries. It is constrained within $1$ and $2$, where $1$ corresponds to linear, sinusoidal--like signals and $2$ is for Gaussian signals covering the whole area. Hence, larger values of HFD indicate greater temporal complexity and irregularity. We compute HFD using the already proposed algorithm \cite{higuchi_approach_1988}.  

For a uniformly sampled timeseries $X$ of length $N$
\[
X(1), X(2), X(3), \ldots, X(N),
\]
we construct a family of $k$ new sub--timeseries $X_k^{m}$ defined as
\[
X_k^{m} =  X(m), X(m+k), X(m+2k), \ldots \,X\left(m+\lfloor\frac{N-m}{k}\rfloor k\right),\]
where $m = 1, 2, \ldots, k$, denotes the initial time offset and $k = 1, 2, \ldots, k_{\text{max}}$ is the time interval between successive points.

For each constructed sub--series $X_k^{m}$, the curve length $L_m(k)$ is calculated as
\begin{equation}
L_m(k) = \frac{1}{k} \cdot 
\frac{(N-1)}{\left\lfloor \frac{N-m}{k} \right\rfloor k}
\sum_{i=1}^{\left\lfloor \frac{N-m}{k} \right\rfloor}
\left| X(m+ik) - X(m+(i-1)k) \right|.
\end{equation}

The mean length for a given $k$ is then obtained by averaging over all $m$:
\begin{equation}
L(k) = \frac{1}{k} \sum_{m=1}^{k} L_m(k).
\end{equation}

The HFD is estimated using the scaling relation
\begin{equation}
L(k) \propto k^{-\rm{HFD}}.
\end{equation}
A linear regression is performed on the $\log L(k)$ versus $\log(1/k)$ plot over the range
$k = 1$ to $k_{\text{max}}$ and the slope of the fit yields HFD.

In this work, we compute HFD using the publicly available Matlab code
\texttt{higfracdim} (\url{https://in.mathworks.com/matlabcentral/fileexchange/124331-higfracdim}).
HFD is highly dependent on $k_{\rm{max}}$. For a timeseries with $\approx1000$ data points, $k_{\rm{max}}$ between 6 and 10 has been recommended \cite{accardo_use_1997, spasic_estimation_2005}. For a larger timeseries, $k_{\rm{max}}$ as $1/100$ the length of timeseries gives reliable estimates \cite{higuchi_relationship_1990}. We follow the same criterion for choosing $k_{\rm{max}}$. We specify it as per its usage in the text.

\subsubsection{Surrogate Data Analysis}

To identify the presence of nonlinearity in signals, we perform the surrogate analysis. We generate surrogate timeseries that preserve the PSD and the amplitude distribution of the original data, while destroying any intrinsic nonlinear structure and compare its HFD with the HFD of original timeseries. If the timeseries is nonlinear, then HFD of surrogates will be distinct than that of original timeseries, else both HFDs will be statistically same. We use the \textit{Iterated Amplitude Adjusted Fourier Transform} (IAAFT) method \cite{schreiber_improved_1996} for generating surrogates. The implementation was carried out using the \textit{Chaotic Systems Toolbox} available through the MathWorks File Exchange (\url{https://in.mathworks.com/matlabcentral/fileexchange/1597-chaotic-systems-toolbox}). 

Briefly, the IAAFT algorithm begins with the generation of phase-randomized Fourier surrogates. In the first step, a new surrogate is obtained by rank-ordering the original timeseries with respect to the previous surrogate, thereby improving the match of the amplitude distribution. In the second step, a new surrogate is then constructed by applying the inverse Fourier transform to the product of the original power-spectrum and the phase of the surrogate obtained in the first step. These two steps are iteratively repeated until the difference between the power-spectrum of the original timeseries and that of the surrogate converges to a minimum. The final IAAFT surrogate, thus, closely reproduces both the amplitude distribution and the power spectrum of the original signal.

Following \cite{prichard_generating_1994}, we generate 39 surrogate realizations for each timeseries, corresponding to a two-sided significance test at the 5\% significance level ($\alpha$). (For a given $\alpha$, number of surrogates (N) = \(2/\alpha - 1\) \cite{schreiber_surrogate_2000}). To quantify nonlinearity, we compute the $z$-score as
\begin{equation}
z-\rm{score} = \frac{|\mathrm{HFD} - \langle \mathrm{HFD}_s \rangle|}{std(\mathrm{HFD}_s)},
\end{equation}
where $\langle \mathrm{HFD}_s \rangle$ denotes the mean HFD of the surrogate ensemble and ${\rm std}(\mathrm{HFD}_s)$ is the corresponding standard deviation. The timeseries is classified as \textit{nonlinear} at the 95\% level if $z-\rm{score} \geq 2$ \cite{prichard_generating_1994}($1.96$ precisely \cite{bevington_data_2003}) for a two-sided test. The adapted $z$-score
has been used as normalized mean sigma deviation ($nmsd$) earlier for surrogate analysis using correlation dimension \cite{harikrishnan_non-subjective_2006, guria_uncovering_2025}. 


\subsubsection{\label{sec_filter}Filtering Data}
In Sec. \ref{Sec:GRS} below, we analyze data observed from the black hole source GRS~1915+105 in order to find natural implications of our results. Generally, observed data are contaminated by noise, which are advisable to filter, as was done in earlier similar analysis \cite{guria_uncovering_2025}.
We filter various temporal classes of GRS~1915+105 lightcurves to reduce noise. We employ the fourth order Chebyshev type I filter that has been previously used in physiological signals like electroencephalogram (EEG) data \citep{aggarwal_changes_2025}. Owing to its steep roll-off in frequency, it is highly effective for suppressing high-frequency noise while preserving the relevant dynamical bandwidth of the signal.

We choose the frequency range for filtering in relevance to the timescales ($k$) probed for HFD analysis. At $k=1$, the sub-timeseries is identical to the original timeseries with sampling frequency $Fs$ containing frequency components from $df$ to $Fs/2$, where $df$ is the frequency resolution. At $k=2$ (with time delay 2), the effective sampling becomes $Fs/2$. Similarly, at $k=k_{\rm max}$, the timeseries is sampled at $Fs/k_{\rm max}$, with frequency information content between $df$ and $Fs/2k_{\rm max}$. This implies that for all $k$ (from $k=1$ to $k_{\rm max}$), the frequency information from $df$ to $Fs/2k_{\rm max}$ is common and is used to provide HFD. Hence, the high frequency of the bandpass filter should be $>>Fs/2k_{\rm max}$ in order to prevent the distortion of temporal information that is being primarily analysed using HFD. The low frequency components ($\sim df$) may represent drifts in timeseries and hence should be removed to bring stationarity. Therefore, we recommend that the observational signals should be bandpass in a range $[\sim df,>>Fs/2k_{\rm max}]$ for denoising before nonlinear analysis.

\subsection{Hurst Index}

The Hurst index is a widely used measure for quantifying \textit{long-range temporal correlations} and \textit{memory effects} in fractal signals \cite{hurst_long-term_1951, mandelbrot_fractals_1997}. Fractal timeseries are commonly classified into two types: fractional Gaussian noise (fGn) and fractional Brownian motion (fBm). These two are interrelated as cumulative summation converts fGn to fBm and fGn can be inversely obtained from fBm via successive differencing. \textit{H} lies between 0 and 1. For fBm, it characterizes the tendency of a signal to either persist ($H>0.5$), anti-persist ($H<0.5$), or behave as an uncorrelated random process ($H=0.5$). However, for fGn, the near neighbour elements are always positively correlated for all $H$ \cite{eke_physiological_2000}. Theoretically, $H$ and FD for a $n$--dimensional system ($n=1$ for a timeseries, $n=2$ for a surface and so on) are related as \cite{orey_gaussian_1970, mandelbrot_fractional_1968}
\begin{equation}
    {\rm FD}+H=n+1.
\end{equation}
Thus, for a timeseries, it results in the following relation
\begin{equation}
    {\rm FD}+H=2.
    \label{Eq_FDH}
\end{equation}
The most common technique to compute $H$ is \textit{Rescaled Range} (R/S) analysis, originally introduced by Harold Edwin Hurst \cite{hurst_long-term_1951}. However, R/S analysis is highly sensitive to heavy-tailed amplitude distributions, trends, and non-stationarities and exhibit finite size biases \cite{weron_estimating_2002,dieker_simulation_2004,barunik_hurst_2010}, often providing a deviated value. Hence, in this work, we estimate the $H$ using the \textit{generalized Hurst exponent} (GHE) formalism \cite{matteo_long-term_2005}, which provides greater robustness for non-Gaussian and intermittent timeseries and shows better convergence properties for moderate-length \cite{matteo_long-term_2005}.

The GHE method is used to compute $H$, based on the $q$th-order moments $K_q(r)$ of the distribution of increments of the timeseries \cite{barabasi_multifractality_1991,matteo_long-term_2005,mandelbrot_fractals_1997}. For the timeseries $X(t)$, the function $K_q(r)$ is defined as
\begin{equation}
K_q(r) = \frac{\left\langle \left| X(t+r) - X(t) \right|^q \right\rangle}
{\left\langle \left| X(t) \right|^q \right\rangle},
\end{equation}
where $r$ is the time lag, varying from $1$ to $r_{\max}$, and $\langle \cdot \rangle$ denotes averaging over time. The parameter $r$ plays a role analogous to $k$ in the HFD calculation.
The function $K_q(r)$ is assumed to follow the scaling relation
\begin{equation}
K_q(r) \sim r^{qH(q)},
\end{equation}
where $H(q)$ is the generalized Hurst exponent. For $q = 1$, $H(1)$ reduces to the classical Hurst exponent $H$.

In this work, we compute $H$ from the first-order moments using a publicly available Matlab implementation of the generalized Hurst exponent algorithm \cite{aste_generalized_2024}, with the default parameter values $q = 1$ and $r_{\rm{max}} = k_{\rm{max}}$.

\subsection{Slope of Power Spectral Density}
Power-law spectral slopes provide a fundamental characterization of how variability power is distributed across temporal or spatial scales in a system. It reflects scale-invariant (self-similar) dynamics, where no characteristic timescale of signal dominates the fluctuations and is a direct signature of correlated transport. Thus, the spectral slope complements temporal-domain based FD, and $H$ by quantifying the frequency-domain manifestation of nonlinear, multiscale dynamics. Theoretically, spectral slope is inversely proportional to FD and, thus, directly varies with $H$ (the exact relation depends on whether the system is fGn or fBm \cite{higuchi_relationship_1990}). Hence, computing it along with FD and $H$ simultaneously helps to gain high confidence on the observed trends across the different simulated conditions.  

We compute PSD using fast Fourier transform (FFT). We calculate the slope of the PSD using Matlab wrapper for \textit{Fitting Oscillations and One Over f} (FOOOF) toolbox \cite{donoghue_parameterizing_2020}. In FOOOF, PSD is modelled as a combination of Gaussian peaks and power-law fall off, therefore, FOOOF is highly convenient for extracting peak and fall-off properties of PSD. The power-law component, also known as `aperiodic component', is obtained as
\begin{equation}
    AP(f) = 10^b f^{-\chi_0},
\end{equation}
where $\chi_0$ is the slope and $b$ is the offset.

We keep the standard FOOOF parameters \cite{aggarwal_slope_2023, aggarwal_changes_2025}. Since the temporal resolution of the HARMPI simulations is $10 \:\: \mathrm{r_g/c}$, this yields the sampling frequency of $0.1 \:\:(\mathrm{r_g/c})^{-1}$. Using Nyquist frequency theorem, the available frequency for analysis is $>0.05 \:\:(\mathrm{r_g/c})^{-1}$. Therefore, we choose a frequency range of $[0.0008, 0.0417] \: (\mathrm{r_g/c})^{-1}$ for slope analysis throughout this work.

\section{\label{Sec_Results}Analyzing timeseries: Simulated data}
Based on above techniques, here we analyze simulated 
timeseries supplemented by observed data. We first consider
time-dependent GRMHD accretion/outflow simulation outcome.
The underlying time evolutions of $\dot{M}$, $\phi$ and 
$\eta$ serve as synthetic timeseries; also serve as the 
proxy of lightcurve for the present purpose. We will have timeseries for MAD and SANE, each of which will have results for different spins of black hole. Further, two different simulation codes are explored to generate conclusive results. However, first we report the results based on HARMPI.

\subsection{Jet analysis}
We consider the outflow efficiency $\eta$ as the proxy for jet dynamics. We compute HFD, $H$ and slope for simulated MAD and SANE disks based on HARMPI, as discussed below.
\subsubsection{Schwarzschild black hole}
\begin{figure*}
    \includegraphics[scale = 0.52]{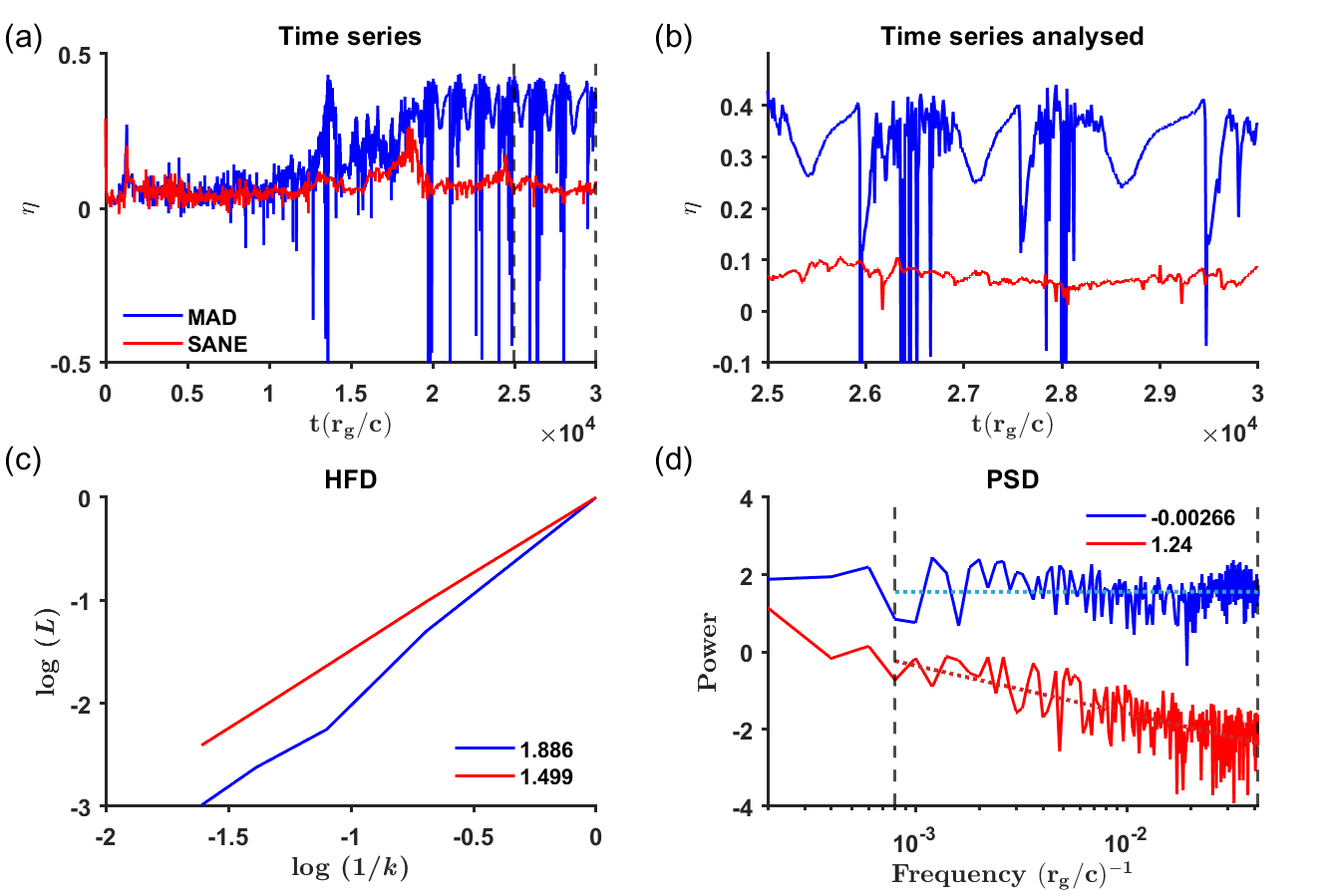}
    \caption{Comparison between MAD and SANE for a Schwarzschild black hole based on HARMPI simulations. (a) $\eta$ timeseries for full simulation period of $30000$ $\mathrm{r_g/c}$. (b)  $\eta$ timeseries used for analysis i.e.  for $25000-30000$ $\mathrm{r_g/c}$. (c) Corresponding log (L) vs log (1/k) curve, used to compute HFD. (We have translated these curves such that they have same value at $k=1$ for better depiction of their slopes). The legends denote the HFD for MAD (blue) and SANE (red). (d) Corresponding PSD with fall-off fitting highlighted by dotted lines. The legends represent slopes computed between 0.0008 and 0.0417 $(\mathrm{r_g/c})^{-1}$ (highlighted by black lines) for MAD (blue) and SANE (red).} 
    \label{fig1_spin0}
\end{figure*}

We begin by considering the simplest case, i.e. spinless black hole described by the Schwarzschild spacetime. Fig. \ref{fig1_spin0}a shows $\eta$ for MAD and SANE for simulated timeseries in time span $t=0 - 30000 \:\mathrm{r_g/c}$. The simulation reach a steady MAD state only after 20000 $\mathrm{r_g/c}$. To ensure stability within the system and stationarity of timeseries, we analyze data from 25000 to 30000 $\mathrm{r_g/c}$ throughout the present work, highlighted in Fig. \ref{fig1_spin0}b. MAD has powerful jets, thereby, possessing high $\eta$ while SANE has weak or no jets, thus, characterized by lower $\eta$. We compute HFD using the slope of log $L$ vs. log ($1/k$) curve with $k_{\rm{max}}=5$ as depicted in Fig. \ref{fig1_spin0}c. MAD has higher HFD than SANE ($\mathrm{HFD}_{\rm{MAD}} = 1.89;\:\mathrm{HFD}_{\rm{SANE}} = 1.50$). Further, $z$-score obtained using $39$ surrogates are $2.35$ and $5.61$ for MAD and SANE respectively. As MAD has HFD closer to 2 with $z$-score just $\gtrsim2$, it suggests exhibiting significant stochastic components. We further validate this by computing PSD slope. The PSDs are depicted in Fig. \ref{fig1_spin0}d  for the corresponding $\eta$ timeseries of Fig. \ref{fig1_spin0}b. The slope is $\sim 0$ for MAD highlighting its stochastic behavior, while it is $1.24$ for SANE indicating it power-law. 
The $H$ for MAD and SANE are $0.04$ and $0.62$ respectively. The resulting sum of HFD and $H$ are $1.93$ and $2.12$ respectively (both close to 2), thus, validating the computation of HFD.

Both MAD and SANE are based on advective accretion flows, effective to explain observed hard X-ray emissions (corroborating with observed jet). An ideal advection dominated accretion flow (ADAF) \cite{Narayan+Yi1994} is expected to be very much deterministic without any cooling \cite{adegoke_correlating_2018}. Also the powerful jets are mostly expected to form by the interplay of the magnetic flux and black hole spin, from Blandford-Znajek process \cite{blandford_electromagnetic_1977}. Here, in absence of spin, only the winds would be formed resulting from Blandford-Payne mechanism \citep{blandford_hydromagnetic_1982, yuan_hot_2014}. Thus, SANE at $a=0$ seems behaving as ADAF with no (or minimal) stochasticity. A MAD flow with $a=0$ and strong magnetic field seems leading to outflows or winds producing additional effects on ADAF as random perturbations (unlike that case of high spin fully developed strong jets; see below). This seems causing HFD of MAD higher. 

\subsubsection{Rotating black hole with different spins}
\begin{figure*}
    \centering
    \includegraphics[scale=0.7]{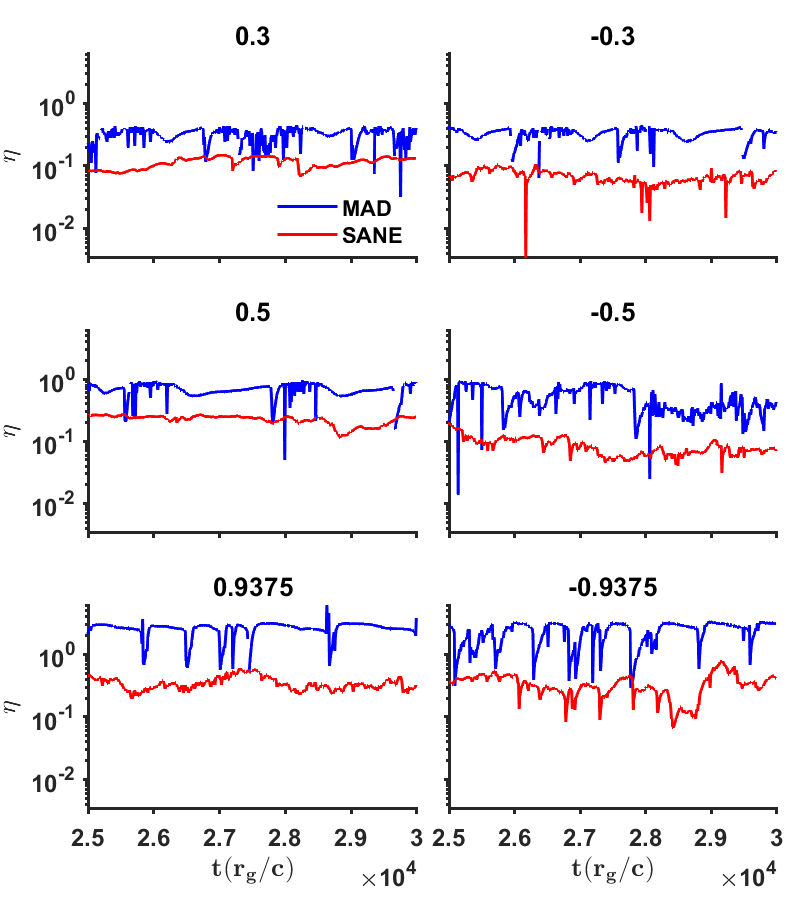}
    \caption{$\eta$ profiles (in log scale) for MAD (blue) and SANE (red) from HARMPI simulations for various spins (left: top to bottom--$a= 0.3, 0.5, 0.9375$, right: top to bottom--$a= -0.3, -0.5, -0.9375$) used for analysis. The discontinuity in MAD timeseries are negative data points that could not be plotted on log scale.}
    \label{fig2_accretionSpin}
\end{figure*}

Next, we observe the variation of the nonlinear characteristics with black hole spin. Fig. \ref{fig2_accretionSpin} shows $\eta$ profiles for MAD and SANE simulations for different spins spanning highly counter-rotating to highly co-rotating black holes for time 25000 to 30000 $\mathrm{r_g/c}$.  $\eta$ for SANE is consistently lower than MAD for all spins. This is expected, due to lesser magnetic flux in SANE compared to MAD \cite{rohan2026}. The jet activity increases in magnitude on considering systems with nonzero spin, both for co- and counter-rotating cases, observed as the upward shift of the timeseries with increasing spin in Fig. \ref{fig2_accretionSpin}. Although the magnitude seems to be similar for co- and counter-rotating cases for a given spin, the fluctuations are more for negative spins as compared to their positive counterparts, except for MAD with $|a|=0.3$. This may be because a small counter-rotation is unable to excel outflow activity overcoming the resistance arising from the inertia of the flow. For the quantitative understanding of these fluctuations, we perform the nonlinear analysis.

\begin{figure*}
    \centering
    \includegraphics[scale=0.5]{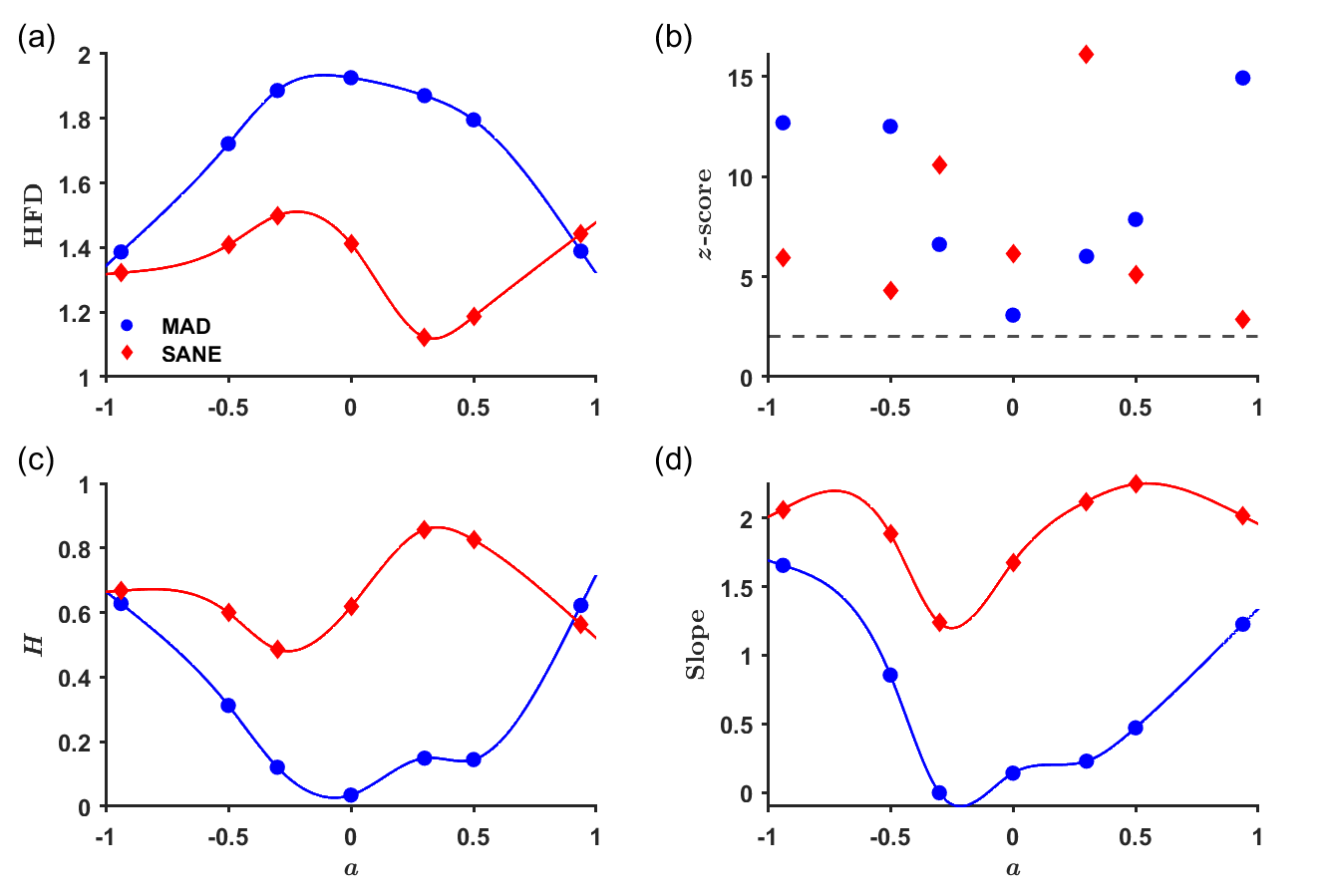}
    \caption{(a) HFD, (b) $z$-score, (c) $H$ and (d) PSD slope for $\eta$ timeseries for MAD and SANE based on HARMPI simulations. The analysed data points are connected using spline fitting in (a), (c) and (d). The dashed line in (b) represent the threshold of $2$, corresponding to 95\% significance level, above which the timeseries could be considered nonlinear. For a reliable nonlinear comparison, an opposite trend to that of HFD between MAD and SANE should be visible across $H$ and slope. As the cross-over at $a=0.9375$ is visible only in HFD and $H$ but not in slope, one cannot comment strongly about the cross-over. }
    \label{fig3_HFD_accretion_spin}
\end{figure*}

The variations of corresponding nonlinear measures are given in Fig. \ref{fig3_HFD_accretion_spin}. HFD (Fig. \ref{fig3_HFD_accretion_spin}a) is consistently higher for MAD than SANE for all the spins except at the extrema ($a=-0.9375$: where both HFDs are very close, and $0.9375$: where HFD of SANE is even higher than MAD). The corresponding $z$-score in Fig. \ref{fig3_HFD_accretion_spin}b shows that the timeseries, except for MAD $a=0$ and SANE $a=0.9375$, are nonlinear, thus, justifying the use of nonlinear analysis for their characterization. Further, the $z$-score for MAD increases with $a$ for both positive and negative cases, indicating an increase in nonlinearity with $|a|$. The higher HFD for MAD than SANE indicates that the jets carry more stochasticity and have lesser long-range temporal correlations than SANE. We supplement this trend based on $H$ as a direct measure of long-range correlations. It has an opposite variation than HFD (see Fig. \ref{fig3_HFD_accretion_spin}c). The average HFD+$H$ for MAD and SANE are $1.99\pm0.04$ and $2.00\pm0.02$ respectively, satisfying the theoretical relation between HFD and $H$ as mentioned in Eq. \ref{Eq_FDH}. The PSD slopes in Fig. \ref{fig3_HFD_accretion_spin}d also follow a similar trend as $H$, indicating more stochasticity for MAD simulations. However, it does not have a crossover at $a=0.9375$ unlike HFD and $H$. At $a=0.9375$, the trend observed from theoretical correlations between HFD and the spectral slope is not seen. 
Moreover, at this spin, the SANE case exhibits a low 
$z$-score, implying that the analysis should not involve a comparison of nonlinear behavior in both accretion regimes, unlike other spins where both MAD and SANE have high $z$-score (Fig. \ref{fig3_HFD_accretion_spin}b). 
In other words, the $a=0.9375$ case effectively contrasts the nonlinear dynamics of the MAD flow with the largely linear variability in the SANE flow.
Hence, one cannot comment on the cross-over behavior with confidence and requires additional verification.

An interesting aspect of the spin dependence is that, while the amplitude of activity, e.g. $\eta$, is symmetric for prograde (co-rotating) and retrograde (counter-rotating) spins and increases monotonically with the magnitude of the spin (see Fig. \ref{fig2_accretionSpin}), the HFD and other nonlinear measures exhibit a non-monotonic variation with spin (see Fig. \ref{fig3_HFD_accretion_spin}). This behavior suggests that the underlying fluctuations are not governed by spin alone, but could result from a complex interplay between black hole rotation and magnetic field dynamics.

We summarize the findings as follows. With the increase 
of spin, MAD develops jet with increasing strength, either via Blandford-Payne \cite{blandford_hydromagnetic_1982} or Blandford-Znajek (for higher spin) \cite{blandford_electromagnetic_1977} mechanism. Outflows in the form of jets and winds increase with the spin of the black hole, but the structure of the magnetic fields also plays an instrumental role in the ejection of these outflows. Magnetic fields in MAD systems are ordered and this structured nature of the fields increases with increase in spin. This consequently leads to powerful outflows from the system especially leading to stronger and more structured jets. Wind outflows thus become dynamically insignificant. Stronger very well-defined jets are expected to supersede all the apparently stochastic disk phenomena appeared at low $a$, making the system nonlinear and more coherent. On the contrary, in SANE systems, weak magnetic fields lack any kind of structure, leading to weaker jets or outflows \citep{pathak_simulating_2025}. Wind outflows then become important in the evolution of the system. The random nature of winds combined with the unstructured nature of magnetic fields leads to a deviation from fractal nature, introducing stochasticity, particularly for higher spin due to lack of proper magnetic field structure, unlike MAD. At a very high $a$, the nonlinearity is almost suppressed in SANE leading to a very low $z$-score. Therefore, while the increase in $a$ leads to more nonlinearity and lesser HFD in MAD in the presence of strong magnetic field, a high $a$ tends to destroy nonlinearity in SANE, resulting in increase in HFD. A steady nonlinear feature appears to be due to the interplay between strong magnetic field and black hole spin. The intermediate spin, e.g. $|a|=0.3$, seems to, however, corroborate with SANE's weaker magnetic fields leading with higher $z$-score.

\subsection{Disk Analysis}
\begin{figure*}
    \centering
    \includegraphics[scale=0.7]{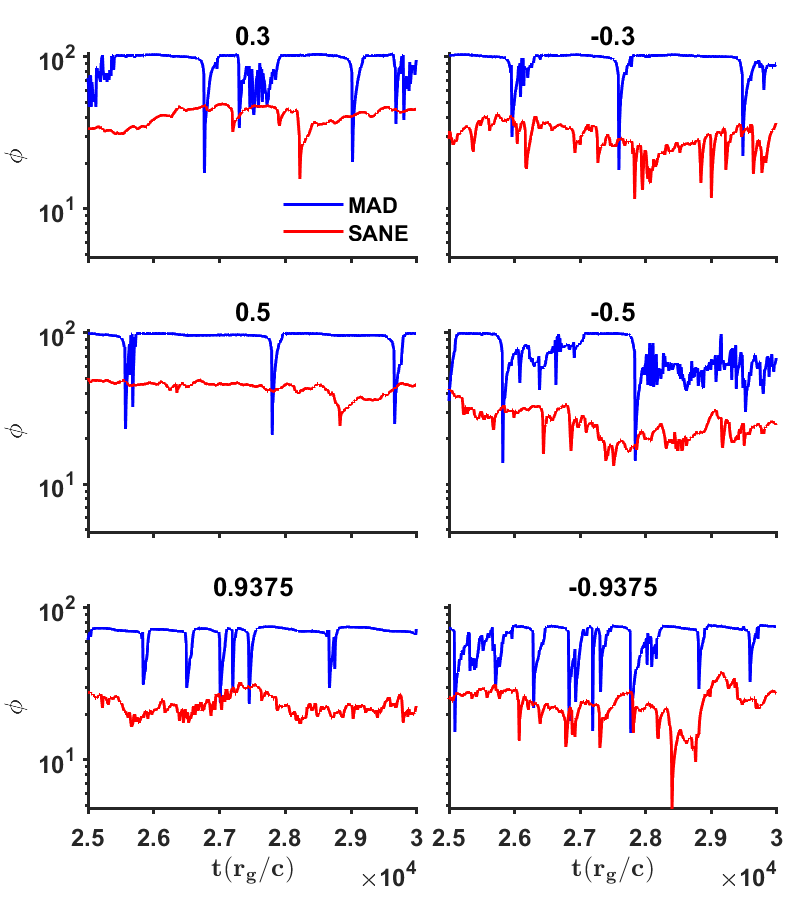}
    \caption{$\phi$ profiles (in log scale) of MAD and SANE using HARMPI simulations for various spins.}
    \label{Fig4_etatimeseries}
\end{figure*}

We consider the normalized flux $\phi$, along with $\dot{M}$, as a proxy for disk dynamics, based on HARMPI simulation. Fig. \ref{Fig4_etatimeseries} shows the timeseries of $\phi$ for MAD and SANE simulations between $25000$ and $30000\:\mathrm{r_g/c}$ for various spins. Although the magnitude of $\phi$ with the increase of spin, for both prograde and retrograde motions, shows the trend opposite to that of $\eta$, the underlying fluctuations follow the same trend as $\eta$ in Fig. \ref{fig2_accretionSpin}. This is because both the disk and jet are governed by the same underlying processes: strong magnetic fields for MAD and MRI for SANE. Thus, the information about the governing processes could be derived from either disk or jet.

\begin{figure*}
    \centering
    \includegraphics[scale=0.5]{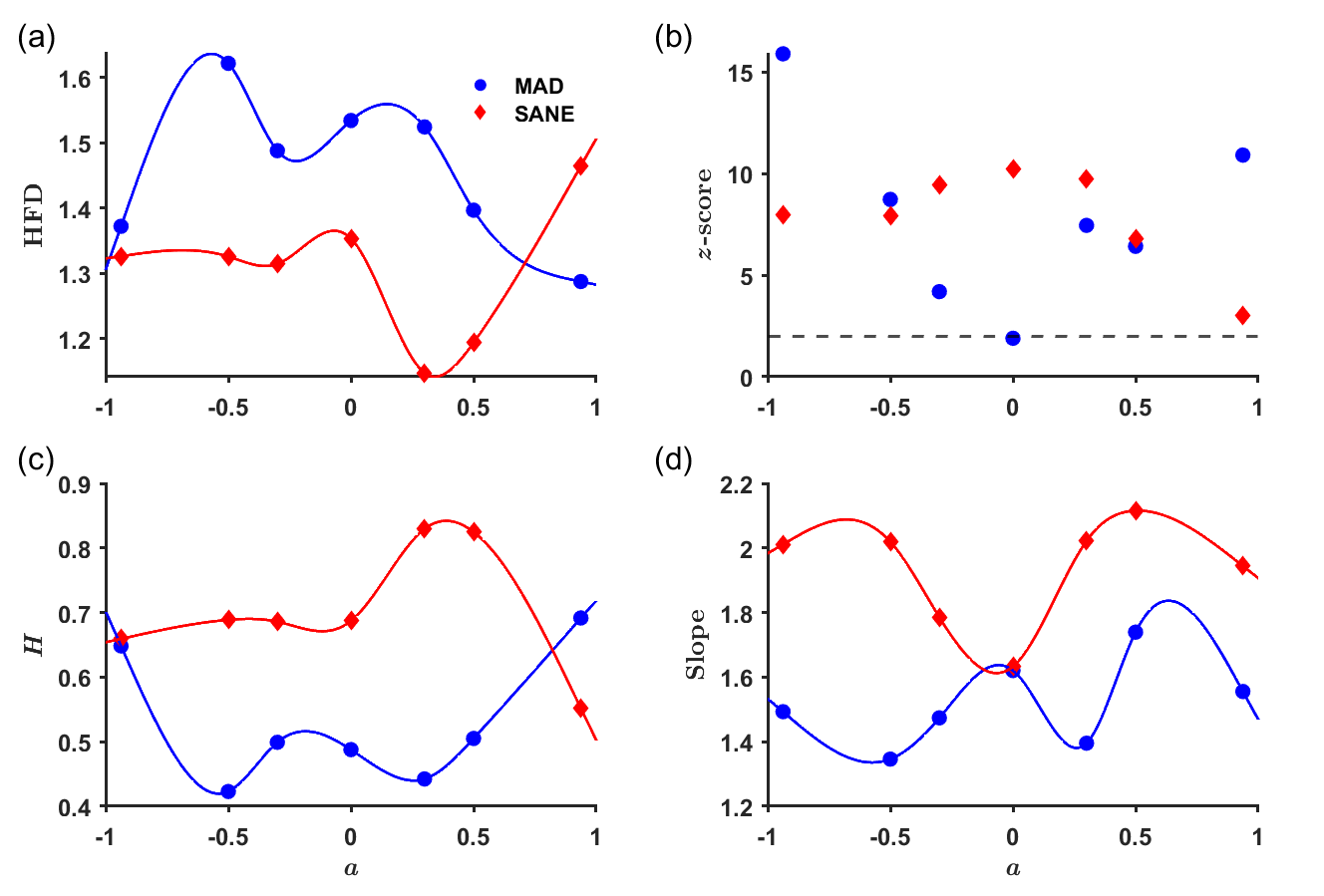}
    \caption{(a) HFD, (b) $z$-score, (c) $H$, and (d) PSD slope, of $\phi$ timeseries for MAD and SANE from HARMPI simulations. Details are same as Fig. \ref{fig3_HFD_accretion_spin}.}
    \label{fig5_etaHFD}
\end{figure*}

Fig. \ref{fig5_etaHFD} represents the variation of nonlinear properties between MAD and SANE for different spins. The trends of HFD, $z$-score, $H$ and slope between MAD and SANE are similar to $\eta$. Even HFD and $H$ for $\phi$ show crossover at $a=0.9375$, and slopes do not. Also the slopes are similar at $a=0$ for MAD and SANE, while HFD and $H$ are distinct. At these spins, 0 and 0.9375, the $z$-score of MAD and SANE $\sim2$ respectively, thus, indicating that the comparison of the nonlinear properties between two disks is not effective. Hence, the trend should be compared with caution at these spins.

All the above disk features remain similar for the $\dot{M}$ based timeseries as well.

\subsection{Validation with BHAC simulations}
\begin{figure*}
    \includegraphics[scale=0.9]{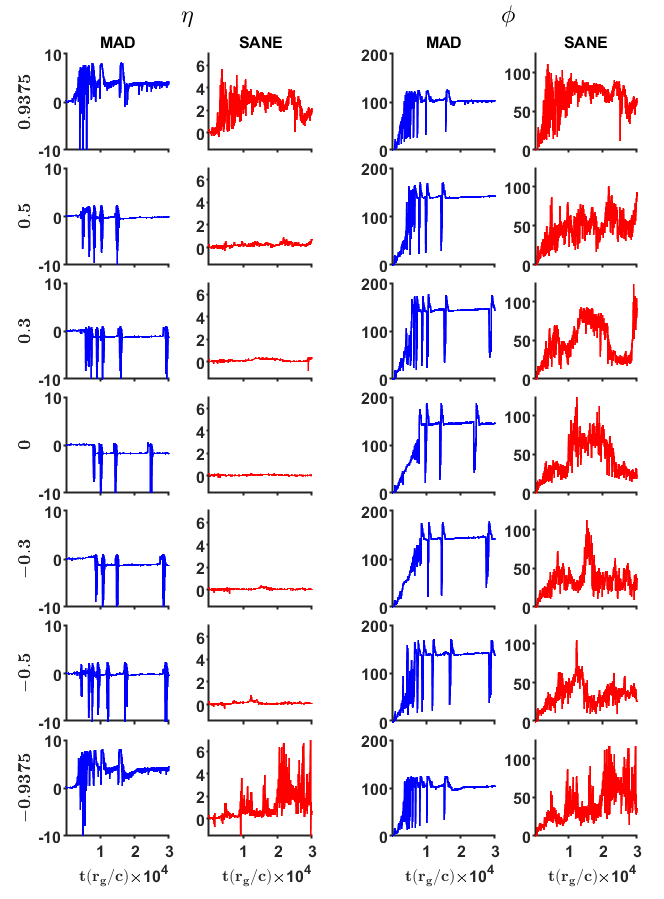}
    \caption{The full timeseries of $\eta$ and $\phi$ for MAD and SANE simulations based on BHAC at various spins mentioned at left (top to bottom--$a = 0.9375, 0.5, 0.3, 0, -0.3, -0.5\:\rm{and}\:-0.9375$) for 0 to 30000 $\rm{rg/c}$. For our analysis, we use the time zone between 25000 to 30000 $\rm{rg/c}$, similar to HARMPI.}
    \label{fig6_BHACTimeSeries}
\end{figure*}

To further validate the variation of nonlinear properties between MAD and SANE, we analyze the timeseries for $\eta$ and $\phi$ obtained using a different simulation setup: BHAC, which also have larger number of data points (with time step 0.1 $\rm{r_g/c}$), given in Fig. \ref{fig6_BHACTimeSeries}. We use the same time window (25000-30000 $\mathrm{r_g/c}$) as HARMPI. As the number of data points is now 50000, we use $k_{\rm{max}}$ as 500 for the HFD computation. 

\begin{figure*}
    \centering
    \includegraphics[scale=0.5]{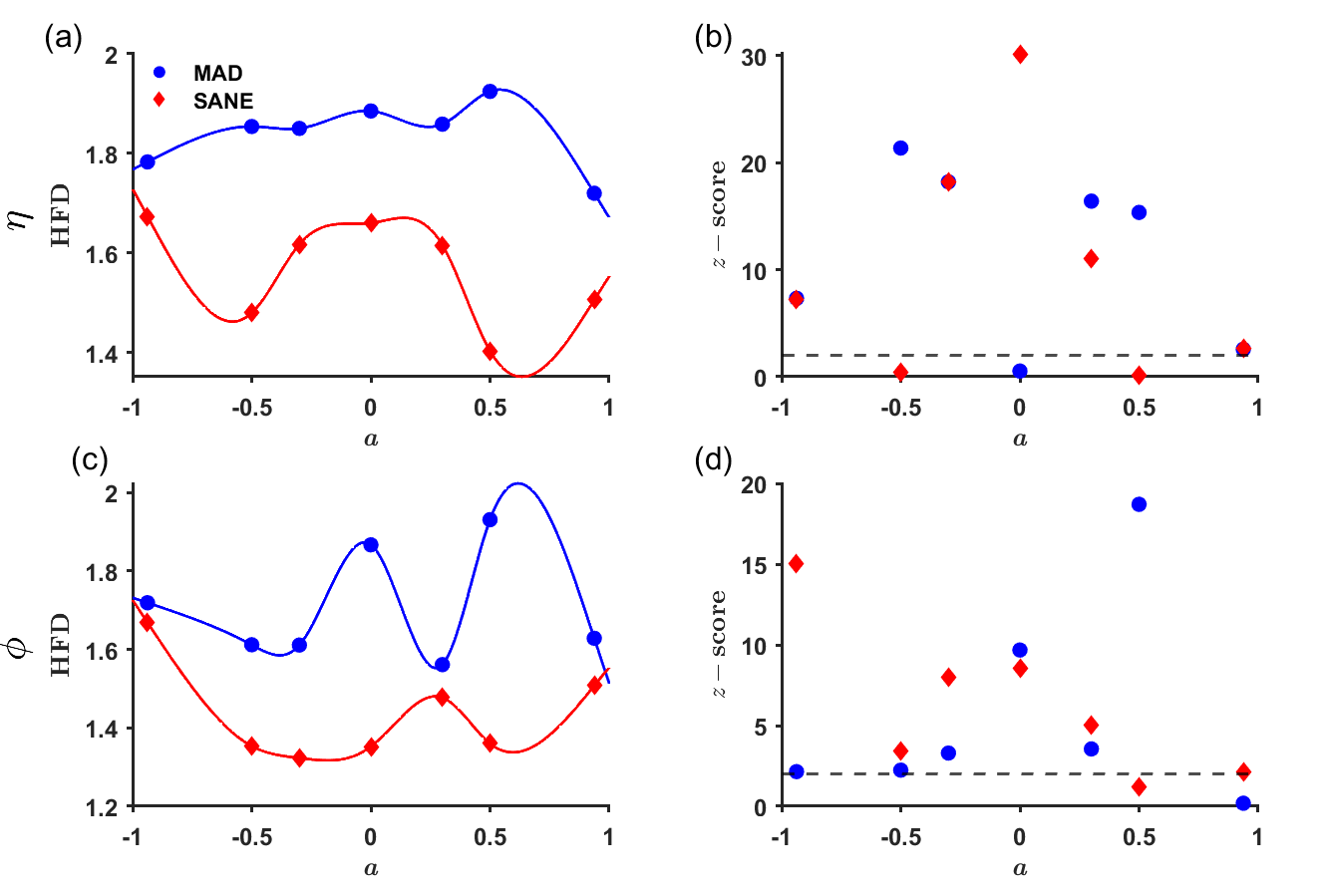}
    \caption{(a,c) HFD, and (c,d) corresponding $z$-score for $\eta $ and $\phi$ respectively for simulations generated using BHAC. The dashed line in (b) and (d) represent the threshold of $2$ above which the timeseries could be considered nonlinear. The data points for HFD in (a) and (c) are connected using spline fitting.}
    \label{fig7_BHACHFD}
\end{figure*}

Similar to HARMPI simulations, the timeseries are mostly symmetric across positive and negative spins and the overall amplitude increases for $\eta$ and decreases for $\phi$, as the magnitude of black hole spin increases. However, here, the simulation system does not have continuous outflows for MAD unlike HARMPI (a work related to code comparison between 
HARMPI and BHAC is underway: Pathak et al.). Rather, the accretion flow in MADs for BHAC features flux eruption events which are of higher period and higher variability than HARMPI. Even so, HFD for MAD is consistently higher than SANE for both $\eta$ and $\phi$ in BHAC as shown in Fig. \ref{fig7_BHACHFD}. For $\eta$, HFD for MAD simulations is similar in the lower spin regime (including 0 and retrograde cases) and decreases at high spins, while for SANE, it decreases with increasing spin magnitude and then increases at extreme spins (see Fig. \ref{fig7_BHACHFD}a). At high retrograde and prograde spins of $-0.9375$ and $0.9375$, respectively, HFD of MAD and SANE come closer but do not exhibit a crossover unlike HARMPI simulations. The corresponding $z$-scores for BHAC simulations in Fig. \ref{fig7_BHACHFD}b indicate that timeseries of either MAD or SANE for multiple spins are linear with their $z\rm{-score} \lesssim2$, thus, making the comparison between MAD and SANE for a given spin ineffective. Here, particularly at $a=0.9375$, both MAD and SANE 
timeseries appear to be close to linear. Thus, at high spin, we cannot confirm the nonlinear properties between MAD and SANE for BHAC. 

For $\phi$, MAD's HFD shows a variable trend with spin as depicted in Fig. \ref{fig7_BHACHFD}c. Also, the inferences about nonlinearity for MAD from the corresponding $z$-score in Fig. \ref{fig7_BHACHFD}d do not match with $z$-scores for $\eta$ in \ref{fig7_BHACHFD}b.
It may be due to the erratic nature of MAD flows in BHAC because of strong flux eruption events that leads to a disconnect between jet and disk dynamics. On the other hand, SANEs' HFD as well as $z$-score trends for $\phi$ (Fig. \ref{fig7_BHACHFD}c and d) and $\eta$ (Fig. \ref{fig7_BHACHFD}a and b) corroborate quite well. This is due to the absence of flux-eruption events in SANEs.



\section{Implications to the observation of GRS 1915+105}\label{Sec:GRS}

\begin{figure*}
    \centering
    \includegraphics[scale=0.6]{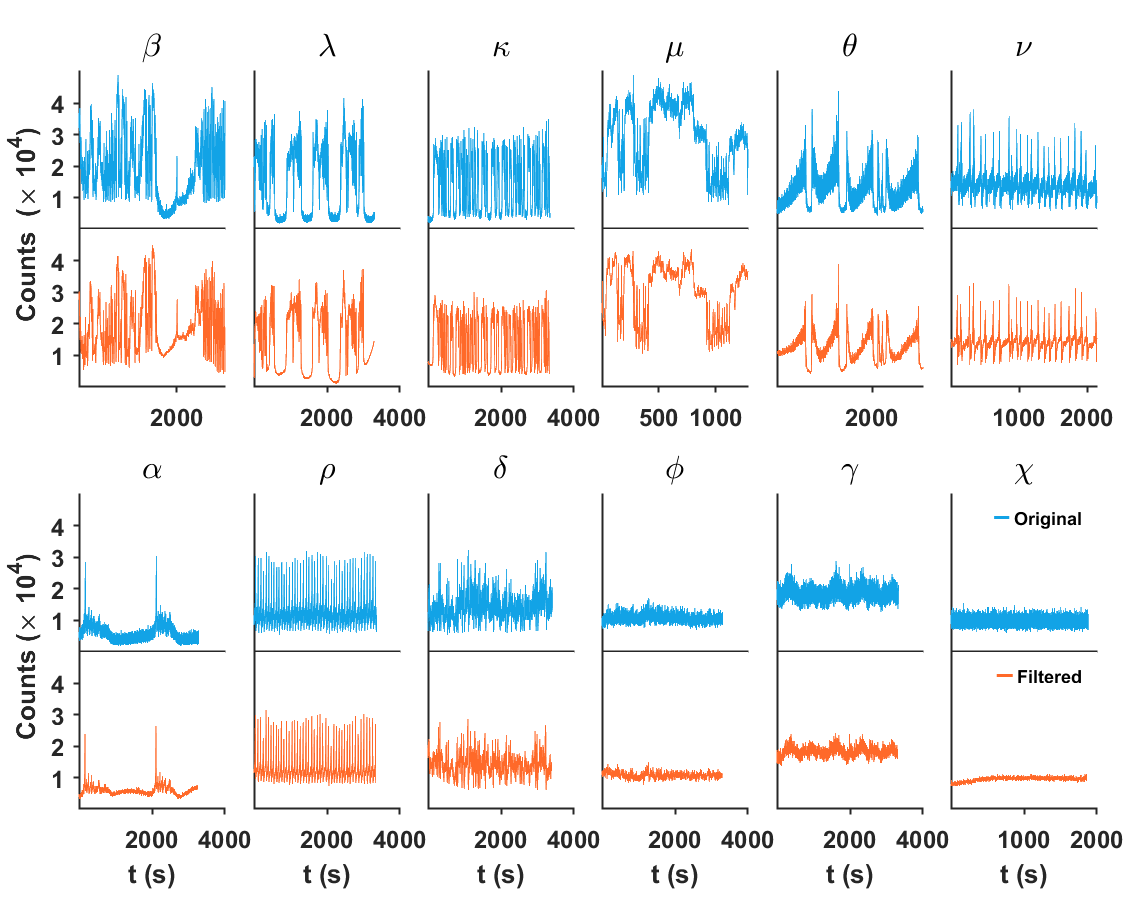}
    \caption{ The original (top) and the filtered (bottom) lightcurves of GRS~1915+105 classes. The filtering is done between $[0.0001, 0.03]Fs$ using fourth order Chebyshev type I filter. The filter effectively reduces high frequency noise. }
    \label{Fig_GRSTimeSeries}
\end{figure*}

\begin{table*}
\centering
\begin{tabular}{l c c c c c c c c c c c}
\hline
ObsID & Class & HFD & $z$-score & Behaviour   & Behaviour  & PL & diskbb & SI & $\chi^2/\nu$ & Count rate & $L$ \\
 & & & &(via HFD) & (via $\mathrm{D_2}$) & &&&&&\\
\hline
10408-01-10-00 & $\beta$   & 1.489 & 8.6  & NL & F  & 52 & 46 & 3.25 & 0.9683 & 3726 & 0.3901 \\
20402-01-37-01 & $\lambda$ & 1.515 & 5.5 & NL & F  & 46 & 54 & 2.96 & 0.9532 & 2902 & 0.3119 \\
20402-01-33-00 & $\kappa$  & 1.348 & 9.2 & NL & F  & 51 & 49 & 2.99 & 1.1740 & 2526 & 0.2720 \\
10408-01-08-00 & $\mu$     & 1.436 & 3.0  & NL & F  & 41 & 56 & 3.26 & 0.9603 & 3891 & 0.4040 \\
20402-01-45-02 & $\theta$  & 1.647 & 10.2  & NL & F  & 88 & 11 & 3.10 & 1.2880 & 3202 & 0.3536 \\
10408-01-40-00 & $\nu$     & 1.558 & 9.5  & NL & F  & 72 & 28 & 2.93 & 1.3570 & 3202 & 0.3498 \\
20187-02-01-00 & $\alpha$  & 1.696 & 4.1  & NL & F  & 77 & 23 & 2.39 & 0.5198 & 1136 & 0.1341 \\
20402-01-03-00 & $\rho$    & 1.483 & 14.0 & NL & LC & 72 & 28 & 2.86 & 0.7870 & 2431 & 0.2659 \\
10408-01-17-00 & $\delta$  & 1.541 & 5.9  & NL & S  & 50 & 48 & 3.43 & 1.4870 & 2601 & 0.2683 \\
10408-01-12-00 & $\phi$    & 1.761 & 0.9  & S  & S  & 34 & 50 & 3.84 & 1.1080 & 2006 & 0.1995 \\
20402-01-56-00 & $\gamma$  & 1.840 & 0.9  & S  & S  & 31 & 60 & 3.78 & 1.1200 & 3521 & 0.3598 \\
10408-01-22-00 & $\chi$    & 1.870 & 8.9  & NL  & S  & 89 & 9  & 3.00 & 1.0700 & 1862 & 0.2107 \\
\hline
\end{tabular}
\caption{Summary of observations and spectral fitting parameters. Columns-- 1. RXTE observational identification id (ObsID). 2. Class name \cite{Belloni2000} 3. HFD. 4. $z$-score using 39 surrogates for filtered timeseries. 11. Behaviour via $z$-score for HFD, (NL: Nonlinear or S: Stochastic). 5. Behavior using correlation dimension \cite{adegoke_correlating_2018} ($\mathrm{D_2}$; F: fractal, LC: limit cycle, S: Stochastic. 6. Power-law component (PL, in percent). 7. Multicolour black body component (diskbb, in percent). 8. PL photon spectral index (SI). 9. Reduced $\chi^2$ ($\chi^2/\nu$) 10. Count rate 11. Total luminosity in 3–25 keV in units of Eddington luminosity for the black hole mass $14\:M_\odot$ \cite{greiner_unusually_2001}. (Columns 5 to 11 are taken from \cite{adegoke_correlating_2018}.}
\label{tab:obs_summary}
\end{table*}

To explore observational implications of our findings of HFD in simulations, we compute HFD for the observed RXTE data of GRS~1915+105 black hole. Based on the lightcurve properties, Belloni and colleagues identified 12 different temporal classes of GRS~1915+105 \cite{Belloni2000}.  We retrieve sample lightcurves for each of the 12 classes from the archival data. 
The IDs and the classes of the data are mentioned in Table \ref{tab:obs_summary} columns 1 and 2. We mostly consider the time resolution 0.1 s for all the lightcurves summing over all energy channels. Further, we derive timeseries from a lightcurve. The timeseries of these classes have mean $28980\pm  2044$ data points (between 13022 and 34051).  
We compute the HFD (with $k_{\rm max}$ set to $1/100$ of the total number of data points) and the $z$-score using the IAAFT method for each class. 
Prior to the HFD analysis, the signals are filtered to reduce noise. As described in Sec.~\ref{sec_filter}, the appropriate filter range should lie within $[\sim df,\ \gg F_s/(2k_{\rm max})]$. 
Given that the minimum number of data points is 13022, the maximum value of $F_s/(2k_{\rm max})$ is $F_s/(2 \times 130) \sim 0.004\,F_s$ (with $k_{\rm max} \sim 1/100$ of the data points). The frequency resolution $df$ is inversely proportional to the total number of data points. Hence, using the maximum data length of 34051, the minimum $df \sim 3 \times 10^{-5}\,F_s$. Taking these limits into account, we filter the timeseries in the range $[0.0001,\,0.03]\,F_s$ using a fourth-order Chebyshev Type I filter. We employ the \textit{filtfilt} function in Matlab, which preserves the phase of the signal. The filtered timeseries are less noisy but not overly smoothed, thereby retaining the intrinsic fluctuations compared to the original timeseries shown in Fig.~\ref{Fig_GRSTimeSeries}.

The HFD and the $z$-score for all the classes of GRS~1915+105, shown in Fig.~\ref{Fig_GRSTimeSeries},  are mentioned in Table \ref{tab:obs_summary} columns 3 and 4. The $z$-score for classes $\phi$ and $\gamma$ are $<2$, indicating their stochastic (S) behavior. The other classes display nonlinear (NL) behavior with $z$-score $>2$, which could be close to fractal and could encompass limit cycles. This classification matches with the classification via correlation dimension \cite{adegoke_correlating_2018}, given in Table \ref{tab:obs_summary} column 5, except $\delta$ and $\chi$. While $\delta$ has been classified as nonlinear by autoencoder based `deviation from stochasticity' (DS) measure \cite{pradeep_measuring_2023}, the nonlinearity is shown to be present in $\chi$ here, which is reported for the first time. Interestingly, the nonlinearity in $\chi$ is detected via HFD even for the unfiltered lightcurve, thereby, ruling out the present nonlinearity as the effect of filtering. Since HFD is a measure of complexity within nonlinear systems, it should not be used to make inference about systems that do not show nonlinearity. Hence, for further analysis and inferences, we consider the 10 nonlinear classes of GRS~1915+105, i.e. all classes except $\phi$ and $\gamma$.

In order to verify whether the MAD systems have indeed higher HFD than SANE, we divide the 10 nonlinear classes of GRS 1915+105 into two clusters, based on their spectral properties as given in Table \ref{tab:obs_summary} columns 6 to 11. For this, we employ \textit{k}-means clustering, an unsupervised machine learning algorithm. We use Matlab inbuilt function \textit{kmeans} for obtaining the two clusters using the spectral properties specified in columns 7-12 in Table \ref{tab:obs_summary}, namely, power-law component (PL), mulicolour black body component (diskbb), PL photon spectral index (SI), reduced $\chi^2$ ($\chi^2/\nu$), count rate and total luminosity in 3–25 keV in units of Eddington luminosity for the black hole mass $14\:M_\odot$ \cite{greiner_unusually_2001}. We visualize and validate the classification using Principal Component Analysis (PCA) (via \textit{pca} function in Matlab). 

\begin{figure*}
    \centering
    \includegraphics[scale=0.52]{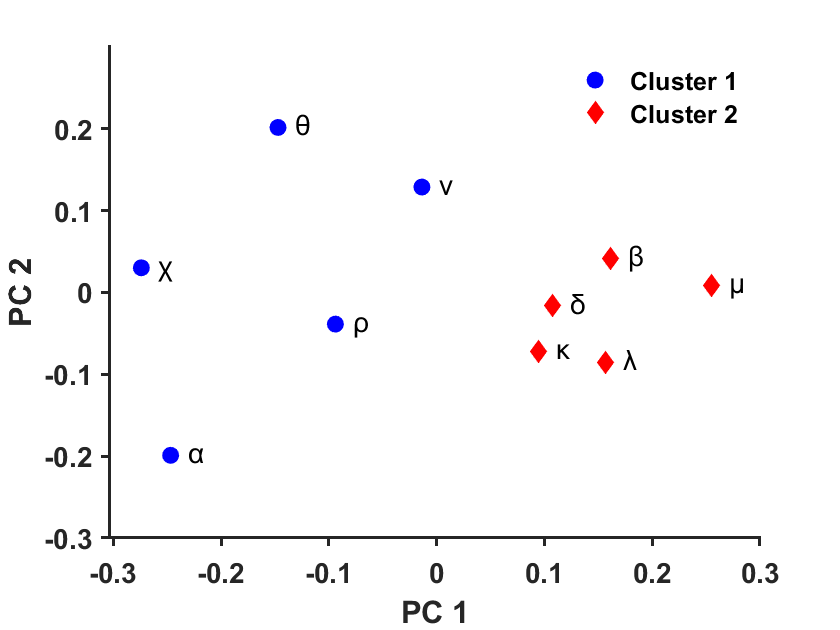}
    \caption{The two principal components based on spectral properties for the ten nonlinear GRS 1915+105 temporal classes. The Clusters 1 and 2 are generated using $k$-means.}
    \label{fig8_GRS}
\end{figure*}

The classes lying in the two clusters, highlighted as blue and red, show clear separability in their first principal component (PC1) as depicted in Fig. \ref{fig8_GRS}. Cluster 1 is primarily dominated by PL spectrum, while cluster 2 classes have significant diskbb spectrum along with PL, as mentioned in Table \ref{tab:cluster_summary}. Notably, Mukhopadhyay and colleagues \cite{adegoke_correlating_2018} classified the majority classes of Cluster 1 ($\theta,\:\nu$ and $\alpha $) as ADAF and most classes of Cluster 2 ($\beta,\:\lambda,\:\kappa$ and $\mu$) as the Keplerian disk. However the classes $\beta,\:\lambda,\:\kappa\:\mathrm{and}\:\mu$ cannot be pure Keplerian, since they have significant (almost $50\%$) PL component. Further, $\rho$ appeared to be between Keplerian and general advective accretion flow (GAAF) and $\chi$ close to GAAF, while $\delta$ a transition between Keplerian and slim disk, according to the previous authors \cite{adegoke_correlating_2018}.

\begin{figure*}
    \centering
    \includegraphics[scale=0.6]{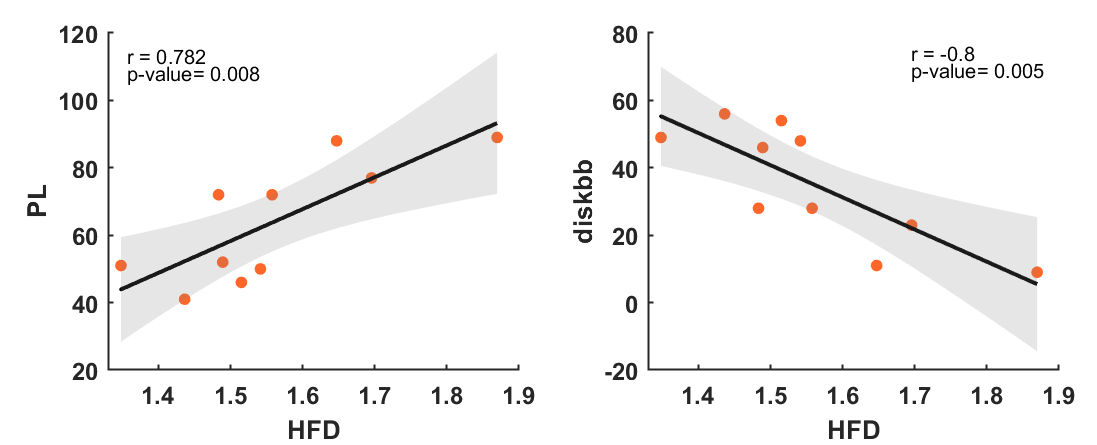}
    \caption{HFD for the filtered lightcurves of ten nonlinear GRS 1915+105 classes vs. their PL (left) and diskbb (right) extracted from Table. \ref{tab:obs_summary}. The classes are represented as orange circles. Regression curve is in black. Shaded region represents confidence interval. The correlation coefficent (r) and p are mentioned at the top. }
    \label{fig9_HFDCorr}
\end{figure*}

We would like to verify how (if) these clusters have properties similar to MAD and SANE. Note that MAD and SANE both are based on ADAFs (or at best GAAFs), while the clusters appear to have advective and Keplerian behaviors based on \cite{adegoke_correlating_2018}. Recent numerical simulations showed that SANE has been geometrically thinner with more angular momentum domination than MAD, making it closer to a Keplerian disk \cite{rohan2026, pathak_simulating_2025}. This suggests that away from black hole, SANE may merge to a pure Keplerian disk. On the other hand, MAD exhibited features of sub-Keplerian, geometrically thicker angular momentum starved, advective accretion flow in the simulations \cite{rohan2026, pathak_simulating_2025}. Therefore, the Cluster 2 is, in fact, SANE--like (SANE+Keplerian), while Cluster 1 is MAD--like (with dominant PL contribution $77\pm3$). Notably, MAD and SANE both being ADAFs will have significant PL component. However, an additional significant diskbb contribution in Cluster 2: $51\pm2$, suggests that the underlying flow contains a good fraction of thermal, Keplerian accretion disk component \cite{1973Sakura_sunyaev,1973Novikov_Thorne}, thereby, making it more SANE--like, not pure SANE. Thus, classes $\theta,\:\nu,\:\alpha \:\rho\: \mathrm{and} \:\chi$ are MAD--like, while classes $\beta,\:\lambda,\:\kappa,\:\mu \: \mathrm{and} \:\delta$ are SANE-like. The average HFD for MAD--like cluster ($1.651\pm0.066$) is higher than SANE--like cluster ($1.466\pm0.034$; t-test, $p = 0.038$), thus, corroborating the simulation results. 
Interestingly, this classification corroborates with the classification done by other authors \cite{rohan2026}, based on variability, X-ray and radio power of various spectral classes of GRS~1915+105 (also Raha, Mukhopadhyay \& Chatterjee, in preparation).
 PL-dominated states correspond to geometrically thick, magnetically dominated flows where energy is released through intermittent fluctuations that destroy long-range correlations and result in larger HFD. In contrast, states with significant diskbb emission correspond to geometrically thiner (compared to the above cases) disks where due to slower infall, the viscous dissipation smoothens out inflow variability, 
 thereby preserving long-range correlations and yielding systematically lower HFD. To measure a direct association between HFD and these spectral components, we compute the correlation of HFD with PL and diskbb components of all nonlinear classes of GRS~1915+105 using Pearson's method \cite{bevington_data_2003}. There is a strong positive and negative correlation of HFD, with PL and diskbb components, respectively, as shown in Fig. \ref{fig9_HFDCorr} (Pearson's correlation coefficient ($r$) of HFD with PL components-- $r = 0.78,\: p = 0.008$; HFD with diskbb components -- $r = -0.8,\: p = 0.005$). This validates that MAD-like cluster with high PL components should have higher HFD than SANE-like cluster wherein the significant diskbb component reduces HFD.

\begin{table*}
\centering
\begin{tabular}{|c| c| c| c| c|c| c| c| c|}
\hline
Cluster & Classes &  PL & diskbb  & SI & $\chi^2/\nu$ & Count rate & $L$ & HFD \\
\hline
1 & $\theta,\:\nu,\:\alpha,\:\rho,\:\chi$  &  80$\pm$3 & 20$\pm$4 & 2.86$\pm$0.11 & 1.0044$\pm$0.1402 & 2367$\pm$356 & 0.2628$\pm$0.0375 & 1.651$\pm$0.066 \\
\hline
2 & $\beta,\:\lambda,\:\kappa,\:\mu,\:\delta$  & 48$\pm$2 &  51$\pm$2 & 3.18$\pm$0.08 & 1.1086$\pm$0.0924 & 3129$\pm$255 & 0.3293$\pm$0.0258 & 1.466$\pm$0.034\\             
\hline
\end{tabular}
\caption{Mean $\pm$ standard error of mean (SEM) for different properties for the two clusters}
\label{tab:cluster_summary}
\end{table*}

\section{\label{Sec_Conclusion}Conclusion}
In this work, we employ nonlinear techniques, namely FD and $H$, along with spectral slope to characterize the simulated jet and disk dynamics for MAD and SANE profiles. We use mainly simulations based on HARMPI to characterize different accretion disks spanning both positive and negative spins. Additionally, we have verified our results using simulations with higher temporal resolution based on BHAC. We have used Higuchi's method to compute FD as it can be used for short and variable 
timeseries efficiently. We perform surrogate analysis to recognize the nonlinearity of 
timeseries. Further, we have obtained $H$ using GHE method over the commonly used R/S method due to its low variance and bias to heavy-tailed distribution as compared to the latter \cite{barunik_hurst_2010, matteo_long-term_2005}. We have shown that using these methods, the sum of the computed FD, i.e. HFD, and $H$ is close to the theoretical value of `2' \cite{orey_gaussian_1970, mandelbrot_self-affine_1985}. We have also computed the slope of power spectral density of these timeseries and showed that the slope follows an opposite trend as compared to the fractal dimension, thus, relating closely with the theoretical relations between FD and slope \cite{higuchi_relationship_1990}. 

Our analysis has revealed systematic differences in the nonlinear properties of MAD and SANE accretion flows across spins encompassing both retrograde and prograde systems. Across both HARMPI and BHAC simulations, MAD systems consistently exhibit higher HFD, lower \textit{H}, and flatter spectral slopes compared to SANE systems for most spin values, indicating that MAD flows are characterized by greater temporal complexity and reduced long-range correlations. This trend holds for both jet dynamics (quantified through outflow efficiency $\eta$) and disk dynamics (quantified through normalized magnetic flux $\phi$).
MAD shows a decreasing HFD with increasing spin magnitude, owing to the increase in collimated jets. On the other hand, HFD of SANE increases with spin, particularly for the prograde cases, due to the increase in both wind and jet power randomly. 
Further, the mean HFD for observational lightcurves from GRS~1915+105 black hole corroborates our simulation results. By grouping the temporal classes of GRS~1915+105 based on spectral properties into MAD-like (PL-dominated) and SANE-like (significant diskbb contribution) clusters using unsupervised machine learning, we have found that the MAD-like cluster shows higher mean HFD than the SANE-like cluster, thus, consistent with the simulation predictions.

The higher HFD and reduced temporal correlations in MAD systems reflect the magnetically-dominated accretion mediated by flux eruptions, whereas the lower HFD and stronger correlations in SANE flows are consistent with more continuous, turbulence-driven quasi-steady accretion governed by weaker magnetic fields. 

The strong positive correlation between HFD and PL spectral component (and negative correlation with diskbb component) in GRS~1915+105 classes suggests that nonlinear timeseries diagnostics can serve as complementary tools to traditional spectral fitting for distinguishing accretion states. This connection between temporal complexity at the horizon and broadband spectral properties provides a pathway for classifying black hole systems and constraining their magnetic field configurations using X-ray variability data \citep{rohan2026}. Notably, while researchers used correlation dimension saturation to label the GRS~1915+105 classes as Fractal or Stochastic with occasional quantitative comparison of the magnitude of correlation dimension among the classes \cite{misra_chaotic_2004, harikrishnan_nonlinear_2011, adegoke_correlating_2018}, here, we have mainly used HFD to understand the complexity of the fractal or nonlinear systems, with higher HFD indicating greater complexity. Hence, we have found that MAD-like classes have higher complexity than SANE-like classes. Thus, HFD provides an advantage of quantification of complexity within fractal system after the identification of systems being nonlinear (fractal) or stochastic without using phase space and multiple embedding dimensions.

Our simulations are 2.5-dimensional (axisymmetric), which may suppress certain three-dimensional (3D) instabilities and turbulent mixing processes that could affect the measured fractal properties. Extending this analysis to fully 3D GRMHD simulations in future would help to assess the role of non-axisymmetric effects. The choice of analysis time window ($25000-30000~\rm{r_g/c}$) was motivated by numerical stability as well as stationarity of timeseries. Different temporal windows or longer simulation durations might reveal additional spin-dependent features or secular trends not captured here, however, any dynamical signature arising from such unstable regimes should be interpreted with caution. Furthermore, owing to the intrinsic stochasticity of the simulations, individual realizations exhibit variations in their detailed temporal behavior, even under identical physical conditions. Nevertheless, the statistical properties remain consistent on average. Therefore, performing present analyses by averaging over multiple realizations with the same simulation set-up would lead to more robust and reliable inferences. 

We have validated the simulation based data using observational data of the black hole source (GRS 1915+105) with 12 temporal classes \citep{Belloni2000,adegoke_correlating_2018}. This has helped to find observational implications of our work. Detailed analysis of additional black hole X-ray binaries and AGNs would be useful to establish the generality of the HFD-magnetic state connection. 
Future work that combines multi-wavelength observations, particularly radio and infrared data that directly trace jet emission, will help clarify how the nonlinear diagnostic of jet dynamics ($\eta$) relates to the observed jet characteristics and to the disk dynamics. Additionally, exploring the relationship between HFD and other accretion disk parameters, such as viscosity, magnetic Prandtl number \citep{Balbus_2008}, and radiative cooling efficiency, could provide deeper physical insight into the origin of fractal variability in magnetized accretion flows.

The present findings have important implications for understanding magnetic regulation in black hole accretion flows and for observational source classification. The ability to distinguish different magnetized accretion states through nonlinear analysis may prove valuable for interpreting observations from current and future X-ray missions, as well as for connecting horizon-scale variability to the large-scale emission structures resolved by the EHT.

\bibliographystyle{apsrev4-1}
\bibliography{references, ref}

\end{document}